\documentclass[conference]{IEEEtran}
\usepackage{amsmath,amssymb,algorithm,algorithmic,amsthm,cite,graphicx,subfigure}
\IEEEtriggeratref{13}

\def\snr{\mathsf{SNR}}
\def\sinr{\mathsf{SINR}}

\begin{document}
\title{On the Robustness of Coordinated Beamforming\\ to Uncoordinated Interference and CSI Uncertainty}

\author{
\IEEEauthorblockN{George~C.~Alexandropoulos$^1$, Paul~Ferrand$^1$, and Constantinos~B.~Papadias$^2$}
\IEEEauthorblockA{$^1$Mathematical and Algorithmic Sciences Lab, France Research Center, Huawei Technologies France, \\20 Quai du Point du Jour, 92100 Boulogne-Billancourt, France}
emails: \{george.alexandropoulos, paul.ferrand\}@huawei.com \\
$^2$B-WiSE Lab, Athens Information Technology, 151 25 Marousi, Athens, Greece\\ email: cpap@ait.gr}

\maketitle
\begin{abstract}
As network deployments become denser, interference arises as a dominant performance degradation factor.
To confront with this trend, Long Term Evolution (LTE) incorporated features aiming at enabling cooperation among different base stations, a technique termed as Coordinated Multi Point (CoMP).
Recent field trial results and theoretical studies of the performance of CoMP schemes revealed, however, that their gains are not as high as initially expected, despite their large coordination overhead.
In this paper, we review recent advanced Coordinated Beamforming (CB) schemes, a special family of CoMP that reduces the coordination overhead through a joint choice of transmit and receive linear filters.
We focus on assessing their resilience to uncoordinated interference and Channel State Information (CSI) imperfections, which both severely limit the performance gains of all CoMP schemes.
We present a simple yet encompassing system model that aims at incorporating different parameters of interest in the relative interference power and CSI errors, and then utilize it for the performance evaluation of the state-of-the-art in CB schemes.
It is shown that blindly applying CB in all system scenarios can indeed be counter-productive.
\end{abstract}

\section{Introduction}\label{sec:Introduction}
Interference from neighboring base stations (BSs) is still one of the most prominent performance degradation factors in cellular networks resulting in outages or performance losses at the cell edges as well as increasing the need for complex handovers. By capitalizing on the wide deployment of multiple antennas---especially at the BS side---and the advances in multi-antenna signal processing techniques, a new approach for interference management, termed as Coordinated Multi-Point (CoMP) \cite{J:Lee_ComMag}, has made its way into mobile communication standards such as Long Term Evolution (LTE).
CoMP is a broad umbrella name for coordination schemes that aim at realizing multi-user communications \cite{Gesbert2010}, i$.$e$.$, sharing the medium among multiple network nodes over space on top of the possible sharing over time and frequency resources.

Focusing on the downlink and considering Joint Transmission (JT) CoMP and in the theoretical limit of infinitely many distributed antennas, one could exactly pinpoint each Mobile Terminal (MT) and ensure that its intended signal adds up at its position while creating no interference for the other MTs in the network. In this case, interference is not only removed, but is actually harnessed and exploited to increase the received signal power at each MT. However, for the the practical implementation of JT CoMP schemes, tight synchronization among the coordinated BSs as well as sharing of Channel State Information (CSI) \emph{and} data for the targeted MTs are necessary. These requirements are one of the downfalls of JT CoMP in practical cellular networks and make its theoretical gains hard to achieve in practice. It was shown in \cite{Irmer2011} that imperfect or outdated CSI has a very large impact on the performance of JT CoMP schemes. The same held when practical radio-frequency components, such as oscillators with phase noise, were considered \cite{Jungnickel2014}. On top of that, it has been recently shown in \cite{J:Lozano_Limits, Mungara2015, J:Kountouris_2015}, that CoMP schemes ignoring the Out-of-the-coordination-Cluster Interference (OCI) see their performances degrade dramatically. As an alternative to JT CoMP for the downlink of cellular networks, the so-called Coordinated Beamforming (CB) approach comes with less stringent coordination requirements \cite{Irmer2011}, while retaining a part of the JT CoMP performance gains. In particular, the practical implementation of CB schemes only shares CSI for the targeted MTs among the coordinated multi-antenna BSs, without any form of user data exchange. The theoretical design of CB schemes with multiple-antenna BSs and MTs has been lately the subject of many research papers \cite{J:Jafar_interference, J:Iterative_Jafar, J:Luo_TSP_IterMMSE, J:Alexandg_Recon2013,Alexandropoulos2016}. Only very recently \cite{Mungara2014} was the spectral efficiency of the CB scheme known as interference alignment (IA) \cite{J:Jafar_interference} studied under time-selective continuous fading channels that are explicitly estimated through pilot symbol observations.

In this paper, we investigate the performance of the CB schemes presented in \cite{J:Jafar_interference, J:Iterative_Jafar, J:Luo_TSP_IterMMSE, J:Alexandg_Recon2013,Alexandropoulos2016} under OCI and CSI imperfections. Following the results \cite{J:Heath_Distributed_Antenna_OCI_2011,J:Heath_Bai_SG_2013} in modeling the power of the received aggregate OCI as a Gamma Random Variable (RV), we present a multi-antenna system model incorporating OCI as an additive Nakagami-$m$ distributed \cite{J:Nakagami} random vector. We also consider imperfect CSI due to errors in pilot symbol assisted estimation or latency. The proposed model allows us to assess the performance of a wide range of practical CB deployments. We conclude on the possible follow-ups to improve the resilience of the latest CB schemes to OCI and CSI uncertainty.

\textit{Notations:} Throughout this paper vectors and matrices are denoted by boldface lowercase letters and boldface capital letters, respectively. The transpose conjugate and the determinant of matrix $\mathbf{A}$ are denoted by $\mathbf{A}^{\rm H}$ and $\det\left(\mathbf{A}\right)$, respectively, $\mathbf{A}^{(n)}$ denotes the $n$th column of $\mathbf{A}$, whereas $[\mathbf{A}]_{i,j}$ and $[\mathbf{a}]_i$ represent the $(i,j)$-element of $\mathbf{A}$ and the $i$-element of $\mathbf{a}$, respectively. In addition, $\mathbf{I}_{n}$ denotes the $n\times n$ identity matrix, while $||\mathbf{A}||_{\rm F}$ stands for the Frobenius norm of $\mathbf{A}$ and $||\mathbf{a}||_2$ denotes the Euclidean norm of $\mathbf{a}$. The expectation operator is denoted as $\mathbb{E}\{\cdot\}$ and the amplitude of a complex number as $|\cdot|$, whereas notation $X\sim\mathcal{C}\mathcal{N}\left(\mu,\sigma^{2}\right)$ represents a RV $X$ following the complex normal distribution with mean $\mu$ and variance $\sigma^{2}$. A diagonal matrix with $\mathbf{a}$ in its main diagonal is denoted by ${\rm diag}\{\mathbf{a}\}$, and $\mathbb{R}$ and $\mathbb{C}$ represent the set of real and complex numbers, respectively.

\section{A System Model\\ Incorporating OCI and CSI Imperfections}\label{sec:Cellular_Model}
We consider a large multi-antenna cellular network from which we single out $B$ BSs, indexed in the set $\mathcal{B}=\{1,2\dots,B\}$, to form a coordination cluster connected through delay-free links to a common network entity. On some time-frequency resource unit, the BS cluster aims at providing service to $B$ MTs indexed in the set $\mathcal{U}=\{1,2,\dots,U\}$ with $U \geq B$. In particular, each BS $b\in\mathcal{B}$ schedules one MT from its associated set of MTs, denoted by $\mathcal{U}_b$, according to a certain scheduling criterion. All sets $\mathcal U_b$ $\forall$$b$ form a partition of the set $\mathcal{U}$ and each BS $b$ is assumed to be equipped with a $n_{\rm T}^{[b]}$-element antenna array, whereas each MT $u\in\mathcal{U}$ has $n_{\rm R}^{[u]}$ antennas. Assuming perfect synchronization of the downlink transmissions within the coordination cluster, the baseband received vector $\mathbf{y}_b\in\mathbb{C}^{n_{\rm R}^{[b]}\times 1}$ at the MT $b\in\mathcal{U}_b$ can be mathematically expressed as
\begin{equation}\label{Eq:Received_Signal}
\mathbf{y}_b \triangleq \mathbf{H}_{b,b}\mathbf{x}_b + \sqrt{\frac{\alpha}{B-1}}\sum_{\ell\in\mathcal{B},\ell\neq b}\mathbf{H}_{b,\ell}\mathbf{x}_\ell + \mathbf{g}_b + \mathbf{n}_b,
\end{equation}
where $\mathbf{g}_b\in\mathbb{C}^{n_{\rm R}^{[b]}\times 1}$ represents the aggregate OCI given by
\begin{equation}\label{Eq:OCI}
\mathbf{g}_b \triangleq \sqrt{\frac{\beta}{|\mathcal{B}'|}}\sum_{i\in\mathcal{B}'}\mathbf{H}_{b,i}\mathbf{x}_i
\end{equation}
with $|\mathcal{B}'|$ denoting the cardinality of the set $\mathcal{B}'$ of interfering multi-antenna BSs that do not belong in the coordination cluster $\mathcal{B}$. In \eqref{Eq:Received_Signal} and \eqref{Eq:OCI}, $\mathbf{H}_{b,k}\in\mathbb{C}^{n_{\rm R}^{[b]}\times n_{\rm T}^{[k]}}$ with $k\in\mathcal{B}$ represents the actual channel matrix between MT $b$ and BS $k$, and $\mathbf{H}_{b,i}\in\mathbb{C}^{n_{\rm R}^{[b]}\times n_{\rm T}^{[i]}}$ with $i\in\mathcal{B}'$ denotes the actual channel matrix between MT $b$ and BS $i$. Furthermore, $\mathbf{x}_j\in\mathbb{C}^{n_{\rm T}^{[j]}\times 1}$ with $j\in\mathcal{B}\cup\mathcal{B}'$ includes the $d_j$ mutually independent and linearly precoded symbols of BS $j$ intended for MT $j$, for which it must hold $d_j\leq\min(n_{\rm T}^{[j]}, n_{\rm R}^{[j]})$ for correct detection to be possible. Without loss of generality, we assume that each BS $i$ not belonging to the coordination cluster, schedules its associated MT $i$ per time-frequency resource unit. We also assume that all deployed BSs in the network are subject to the total power constraint $P$, therefore, it holds that $\mathbb{E}\{\|\mathbf{x}_j\|_2^2\}\leq P$, and the vector $\mathbf{n}_b\in\mathbb{C}^{n_{\rm R}^{[b]}\times 1}$ represents the zero-mean complex additive white Gaussian noise (AWGN) having covariance matrix $N_0\mathbf{I}_{n_{\rm R}^{[b]}}$. Finally, $\alpha,\beta\in[0,1]$ in \eqref{Eq:Received_Signal} and \eqref{Eq:OCI} are used for modeling the relative strength of the received aggregate Intra-Cluster Interference (ICI) (i$.$e$.$, of the $2$nd term in \eqref{Eq:Received_Signal}) and OCI (i$.$e$.$, of the $3$rd term in \eqref{Eq:Received_Signal}) as will be described in the sequel. It is noted that, in general, $\alpha$ and $\beta$ can take values greater than $1$. As a sanity requirement, here we assume appropriate MT associations to BSs that restrict the values of $\alpha$ and $\beta$ in $[0,1]$.

Each actual channel matrix $\mathbf{H}_{b,j}\in\mathbb{C}^{n_{\rm R}^{[b]}\times n_{\rm T}^{[j]}}$ between MT $b$ and any BS $j$ is assumed to include independent elements each modeled as $[\mathbf{H}_{b,j}]_{p,q}\sim\mathcal{CN}(0,(n_{\rm T}^{[j]}n_{\rm R}^{[b]})^{-1})$ with $p=1,2,\ldots,n_{\rm T}^{[j]}$ and $q=1,2,\ldots,n_{\rm R}^{[b]}$. From these channels, the matrices $\mathbf{H}_{b,k}$ $\forall$$k$ are assumed to be estimated sequentially by each MT $b$ using minimum mean squared error (MMSE) channel estimation with $N_{\rm p}$ pilot symbols, and then, fed back through an error-free channel to the common network entity the BSs belonging to the cluster are attached to. More specifically, the estimated and actual channels are related as $\hat{\mathbf{H}}_{b,k}\triangleq\mathbf{H}_{b,k}+\mathbf{E}_{b,k}$, where $\mathbf{E}_{b,k}\in\mathbb{C}^{n_{\rm R}^{[b]}\times n_{\rm T}^{[k]}}$ represents the MMSE estimation error matrix with entries distributed as $[\mathbf{E}_{b,k}]_{i,j}\sim\mathcal{CN}\left(0,(1+\rho\snr)^{-1}\right)$ with $\rho=N_{\rm p}/n_{\rm T}^{[k]}$ and $\snr=P/N_0$ \cite{C:Marzetta_PSAM}. It is noted that, in \cite{Mungara2014}, different forms of $\mathbf{E}_{b,k}$ are presented modeling various realistic imperfections, such as CSI estimation errors and analog CSI feedback. Using the latter assumptions for the system model in \eqref{Eq:Received_Signal}, the power of the intended channel at each MT $b$ and the power of the aggregate ICI (i$.$e$.$, of the $2$nd term in \eqref{Eq:Received_Signal}) are related as 
\begin{equation}\label{Eq:Derivation_alpha}
    \frac{\alpha(B-1)^{-1}\sum_{\ell\in\mathcal{B},\ell\neq b}\mathbb{E}\{\|\mathbf{H}_{b,\ell}\|_{\rm F}^2\}}{\mathbb{E}\{\|\mathbf{H}_{b,b}\|_{\rm F}^2\}} = \alpha.
\end{equation}
For example, $\alpha=1$ models cases where MTs are at similar distances from all BSs in the cluster; under these cases intended channels and ICI are of almost equal powers. The latter value for $\alpha$ is well suited for modeling scenarios according to which MTs are located at the edges of the separate cells (i$.$e$.$, the coordination cluster center), where coordinated transmission is expected to provide its highest gain. Note that \eqref{Eq:Derivation_alpha} holds also for estimated channel matrices (i$.$e$.$, for $\hat{\mathbf{H}}_{b,k}$ $\forall$$k\in\mathcal{B}$). Moreover, the relation between the power of the estimated intended channel at each MT $b$ and the power of the aggregate OCI (i$.$e$.$, of the $3$rd term in \eqref{Eq:Received_Signal}) is given by
\begin{equation}\label{Eq:Derivation_beta}
    \frac{\beta|\mathcal{B}'|^{-1}\sum_{i\in\mathcal{B}'}\mathbb{E}\{\|\mathbf{H}_{b,i}\|_{\rm F}^2\}}{\mathbb{E}\{\|\hat{\mathbf{H}}_{b,b}\|_{\rm F}^2\}} = \frac{\beta}{n_{\rm T}^{[b]}n_{\rm R}^{[b]}(1+\rho\snr)^{-1}+1}.
\end{equation}
It can be seen from \eqref{Eq:Derivation_beta} that for fixed $\snr$, as $N_{\rm p}\rightarrow\infty$ (approaching perfect channel estimation) then $\rho\rightarrow\infty$, and as a result the power of the aggregate OCI at MT $b$ tends to be $\beta$ times the power of its estimated intended channel. The same happens with a fixed $\rho$ value and $\snr\rightarrow\infty$. In general, parameter $\beta$ indicates the effectiveness of BS clustering \cite{C:Papadogiannis_2012} for coordinated transmission. Low values of $\beta$ indicate that most of the interfering BSs for a specific MT have been included within the coordination cluster, the opposite happens for high $\beta$ values.

The elements of the aggregate OCI vector $\mathbf{g}_b$ in \eqref{Eq:OCI} $\forall$$b$ are complex RVs whose amplitudes are modeled in this work as independent and identically distributed (IID) Nakagami RVs \cite{J:Nakagami} with finite shape parameter $m\geq0.5$ and finite $\Omega=\beta P/n_{\rm R}^{[b]}$. It follows that their squared amplitudes are IID Gamma RVs with the following shape and scale parameters   
\begin{equation}\label{Eq:Model_OCI}
|[\mathbf{g}_b]_q|^2 \sim {\rm GAMMA}\left(m,\frac{\beta P}{m n_{\rm R}^{[b]}}\right),\,\forall q=1,2,\ldots,n_{\rm R}^{[b]}.
\end{equation} 
It is trivial to conclude from \eqref{Eq:Model_OCI} that $\mathbb{E}\{|[\mathbf{g}_b]_q|^2\}=\beta P/n_{\rm R}^{[b]}$ and $\mathbb{E}\{\|\mathbf{g}_{b}\|_2^2\}=\beta P$. For the special of $m=1$, \eqref{Eq:Model_OCI} simplifies to the exponential distribution, which implies that the each element of $\mathbf{g}_b$ can be a zero mean complex Gaussian RV with variance $\Omega$. The latter modeling of the OCI's vector elements was used in the analysis of \cite{J:Lozano_Limits}, where it was considered that OCI is made up of a large number of interference terms, and thus can be effectively approximated as Gaussian. For general values of $m\geq0.5$ in \eqref{Eq:Model_OCI}, the adopted model resembles that of \cite{J:Heath_Distributed_Antenna_OCI_2011, J:Heath_Bai_SG_2013, J:Kountouris_2015}, where the amplitude distribution of each element of $\mathbf{g}_b$, computed using both the hexagonal grid and stochastic geometry models, was approximated by a Gamma RV through the moment matching approach. Starting from \eqref{Eq:Received_Signal} and considering the OCI model of \eqref{Eq:Model_OCI}, it can be easily concluded that the received power of the desired signal at MT $b$ with channel estimation and that of its received aggregate OCI are related as in the left-hand side of \eqref{Eq:Derivation_beta}, yielding 
\begin{equation}\label{Eq:Final_Derivation}
    \frac{\mathbb{E}\{\|\mathbf{g}_{b}\|_2^2\}}{\mathbb{E}\{\|\hat{\mathbf{H}}_{b,b}\mathbf{x}_{b}\|_2^2\}} \geq \frac{\beta}{n_{\rm T}^{[b]}n_{\rm R}^{[b]}(1+\rho\snr)^{-1}+1}.
\end{equation}

\section{Downlink Coordinated Beamforming}\label{sec:CB_Schemes}
In this section, the system model of Section~\ref{sec:Cellular_Model} is first employed to a simplistic cellular network in order to demonstrate the theoretical gains of JT CoMP and CB schemes over representative non-coordinated ones as well as to compare JT CoMP and CB under different levels of ICI and OCI. Then, we briefly describe the most representative CB schemes to be compared under the considered system model in the following section with the simulation results, and introduce the metric for the performance comparisons.

\subsection{Theoretical Gains of BS Coordination}\label{sec:Example}
\begin{figure}[t!] 
\centering
\includegraphics[width=3.45in]{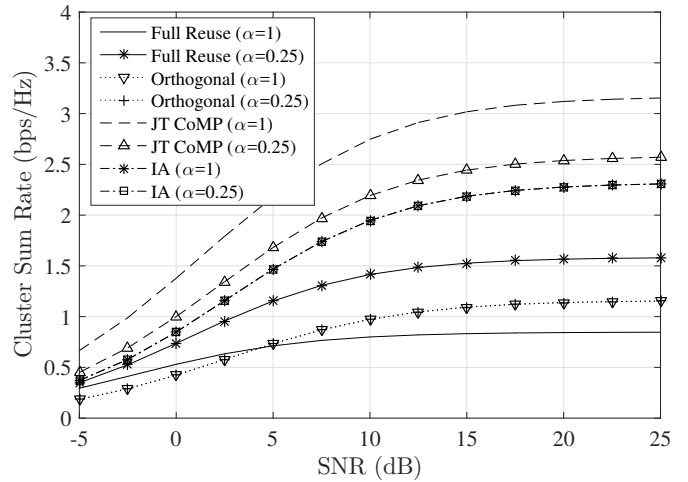} 
\caption{Theoretical sum rate gains of coordination with $\beta=0.25$, perfect CSI, and two different values of $\alpha$. The network sum rate with full reuse of the available time-frequency resources as well as with orthogonal transmissions is also sketched.}
\label{Fig:Comparisons_Toy_a}
\end{figure}

We consider a cluster of $B=2$ multi-antenna BSs, which is a part of a large cellular network. Both BSs in the cluster coordinate their transmissions to serve a total of $2$ MTs in one time-frequency resource unit. One MT is associated to the one BS and the other MT to the other BS. Capitalizing on the system model of Section~\ref{sec:Cellular_Model} for the case of perfect CSI availability, and using the classical bounds for the individual MT rates in multiple-input single-output IFCs \cite{Gesbert2010}, it holds that: 
\begin{itemize}
	\item With full reuse of time-frequency resources, each MT is subject to interference from every BS not associated with and its rate is upper bounded as $\log_2[1+\snr/(\alpha\snr+\beta\snr+1)]$;
	\item With orthogonal allocation of the available time-frequency resources, ICI is absent but the prelog factor $0.5$ appears on each individual MT rate, yielding the value $0.5\log_2(1+\snr/(\beta\snr +1)]$;
	\item With the CB scheme based on interference alignment (IA) \cite{J:Jafar_interference}, ICI can be nulled and the individual MT rate becomes $\log_2[1+\snr/(\beta\snr+1)]$; and
	\item With ideal JT CoMP, the interference power actually boosts the intended signals and the individual MT rate is given by $\log_2[1+(1+\alpha)\snr/(\beta\snr+1)]$.
\end{itemize} 
The latter rates for each of the two individual MTs are added and then depicted in Fig$.$~\ref{Fig:Comparisons_Toy_a} for the case of OCI being $6$ dB lower than that of the intended signal (i$.$e$.$, $\beta=0.25$) and for two different values of $\alpha$ that reveal the relative ICI power. As expected, both coordinated schemes provide substantial gains compared with full reuse and orthogonal transmission when the $\snr$ increases and the network operates in the interference-limited regime. As $\alpha$ approaches $0$ the gain of JT CoMP over IA decreases. For example, for $\snr=15$ dB and $\alpha=1$, IA results in a nearly $100\%$ gain over orthogonal transmission, while this gain becomes nearly $180\%$ for JT CoMP. When $\alpha$ decreases, the latter gain of IA remains the same, whereas that of JT CoMP decreases to nearly $110\%$. This example illustrates that, in many cases of interest, a large part of the gain from coordinated schemes comes more from the removal of interference from the signal of interest rather than from stacking the powers of multiple transmitting points. It is also noted that, when considering practical implementation issues in achieving JT CoMP, the bonus of full coordination becomes even lower, since JT CoMP is more afflicted by degraded CSI and dirty RF than CB \cite{Jungnickel2014}.

\subsection{CB Schemes}\label{sec:Full_Coordination}
The following CB schemes are considered for performance comparisons under the system model presented in Section~\ref{sec:Cellular_Model}: \textit{i}) IA \cite{J:Jafar_interference} that aims at aligning and then nulling interference at each MT belonging in the BS coordination cluster; \textit{ii}) the maximum signal-to-interference-plus-noise ratio (SINR) scheme \cite{J:Iterative_Jafar} that targets at maximizing the received SINR of each transmitted information stream within the cluster; \textit{iii}) the weighted MMSE (WMMSE) scheme \cite{J:Luo_TSP_IterMMSE} that minimizes a system-wide MMSE metric; and \textit{iv}) the Reconfigurable scheme of \cite{J:Alexandg_Recon2013}. The latter CB scheme designs the receive combining matrices at MTs from the minimization of a system-wide MMSE metric and the transmit precoding matrices at BSs are composed of two parts: the one part aims at optimizing the same metric with the receive combiners, while the other part targets at maximizing each individual MT rate. The above CB schemes are more effectively realized through centralized implementations with full CSI exchange among the BSs belonging in the coordination cluster. It is noted, however, that for the maximum SINR, WMMSE, and Reconfigurable schemes, which are iterative as opposed to IA that is not, distributed versions are also available, where explicit CSI exchange among coordinated BSs is avoided. In addition, the latter CB schemes are linear, which means that each BS $b\in\mathcal{B}$ designs its transmitted vector $\mathbf{x}_b$ as $\mathbf{x}_b=\mathbf{V}_b\mathbf{P}_b^{1/2}\mathbf{s}_b$, where $\mathbf{V}_b\in\mathbb{C}^{n_{\rm T}^{[b]}\times d_b}$ represents the precoding matrix, for which we assume that $||\mathbf{V}_b^{(\kappa)}||_2\triangleq1$ $\forall$~$\kappa=1,2,\ldots,d_b$, $\mathbf{s}_b\in\mathbb{C}^{d_b\times1}$ is the information stream vector with $\mathbb{E}\{\mathbf{s}_b\mathbf{s}_b^{\rm H}\}=\mathbf{I}_{d_b}$, and $\mathbf{P}_b\triangleq{\rm diag}\{[P_1^{(b)}\,P_2^{(b)}\,\ldots\,P_{d_b}^{(b)}]\}\in\mathbb{R}_+^{d_b\times d_b}$ with $P_\kappa^{(b)}$ denoting the power allocated to the $\kappa$th information stream. Upon the signal reception given by \eqref{Eq:Received_Signal}, each MT $b$ estimates its desired transmitted symbols using a combining matrix $\mathbf{U}_b\in\mathbb{C}^{n_{\rm R}^{[b]}\times d_b}$, yielding the estimated information stream vector $\hat{\mathbf{s}}_b \triangleq \mathbf U_b \mathbf y_b$. Substituting the linear precoding and combining matrices obtained from any of the aforementioned CB schemes into \eqref{Eq:Received_Signal} yields 
\begin{equation}\label{Eq:Estimated_Symbols}
\begin{split}
\hat{\mathbf{s}}_b =& \mathbf{U}_b^{\rm H}\mathbf{H}_{b,b}\mathbf{V}_b\mathbf{P}_b^{1/2}\mathbf{s}_b + \sqrt{\frac{\alpha}{B-1}}
\\& \times\sum_{\ell\in\mathcal{B},\ell\neq b}\mathbf{U}_b^{\rm H}\mathbf{H}_{b,\ell}\mathbf{V}_\ell\mathbf{P}_\ell^{1/2}\mathbf{s}_\ell + \mathbf{U}_b^{\rm H}\left(\mathbf{g}_b + \mathbf{n}_b\right).
\end{split}
\end{equation}

By inspecting \eqref{Eq:Estimated_Symbols}, it can be easily concluded that the received SINR of each $\kappa$th information bearing stream at each MT $b$ can be expressed as
\begin{equation}\label{Eq:SINR_at_symbol_kappa}
\sinr_{b,\kappa} \triangleq \frac{P_\kappa^{(b)}|[\mathbf{U}_b^{\rm H}\mathbf{H}_{b,b}\mathbf{V}_b]_{\kappa,\kappa}|^2}{I_b+I_{\rm ICI}+|[\mathbf{U}_b^{\rm H}(\mathbf{g}_b+\mathbf{n}_b)]_\kappa|^2},
\end{equation}
where $I_b$, $I_{\rm ICI}\in\mathbb{R}_+$ represent the inter-stream interference from the other $d_b-1$ streams intended for MT $b$, and the interference from the streams belonging in the ICI, respectively, which can be expressed as  
\begin{equation}\label{Eq:Streams_b}
I_b \triangleq \sum_{\lambda=1, \lambda\neq\kappa}^{d_b}P_\lambda^{(b)}|[\mathbf{U}_b^{\rm H}\mathbf{H}_{b,b}\mathbf{V}_b]_{\kappa,\lambda}|^2,
\end{equation}
\begin{equation}\label{Eq:Streams_ICI}
I_{\rm ICI} \triangleq \sum_{\ell\in\mathcal{B},\ell\neq b}\sum_{\mu=1}^{d_\ell}
P_\mu^{(\ell)}|[\mathbf{U}_b^{\rm H}\mathbf{H}_{b,\ell}\mathbf{V}_\ell]_{\kappa,\mu}|^2.
\end{equation}

\subsection{Cluster Sum Rate Performance}\label{sec:Performance_Metrics}
Substituting the linear precoding and combining matrices obtained from any of the CB schemes described in Section~\ref{sec:Full_Coordination} into \eqref{Eq:Received_Signal}, the average achievable sum rate with downlink CB for the coordination cluster is given by
\begin{equation}\label{Eq:Ergodic_SumRate}
R \leq \sum_{b=1}^B\mathbb{E}_{\mathbf{H}}\left\{\log_2\left[\det\left(\mathbf{I}_{n_{\rm R}^{[b]}}+\mathbf{H}_{b,b}\mathbf{S}_b\mathbf{H}_{b,b}^{\rm H}\mathbf{Q}_b^{-1}\right)\right]\right\},
\end{equation}
where notation $\mathbb{E}_{\mathbf{H}}\{\cdot\}$ represents the expectation over all $\mathbf{H}_{b,b}$'s, while $\mathbf{S}_b=\mathbf{V}_b\mathbf{P}_b\mathbf{V}_b^{\rm H}$ $\forall$$b$ denotes the covariance matrices of the transmitted signals and $\mathbf{Q}_b\in\mathbb{C}^{n_{\rm R}^{[b]}\times n_{\rm R}^{[b]}}$ are the ICI plus OCI plus noise covariance matrices, which can be computed as
\begin{equation}\label{Eq:Q_Matrix}
\mathbf{Q}_b = \frac{\alpha}{B-1}\sum_{\ell\in\mathcal{B},\ell\neq b}\mathbf{H}_{b,\ell}\mathbf{S}_\ell\mathbf{H}_{b,\ell}^{\rm H} + \left(\frac{\beta P}{n_{\rm R}^{[b]}}+N_0\right)\mathbf{I}_{n_{\rm R}^{[b]}}.
\end{equation}  
Note that in the computation of the average achievable sum rate in \eqref{Eq:Ergodic_SumRate} we do not include the $\mathbf{U}_b$'s $\forall$$b\in\mathcal{B}$ designed by each of the considered CB schemes.

\begin{figure}[t!] 
\centering
\includegraphics[width=3.45in]{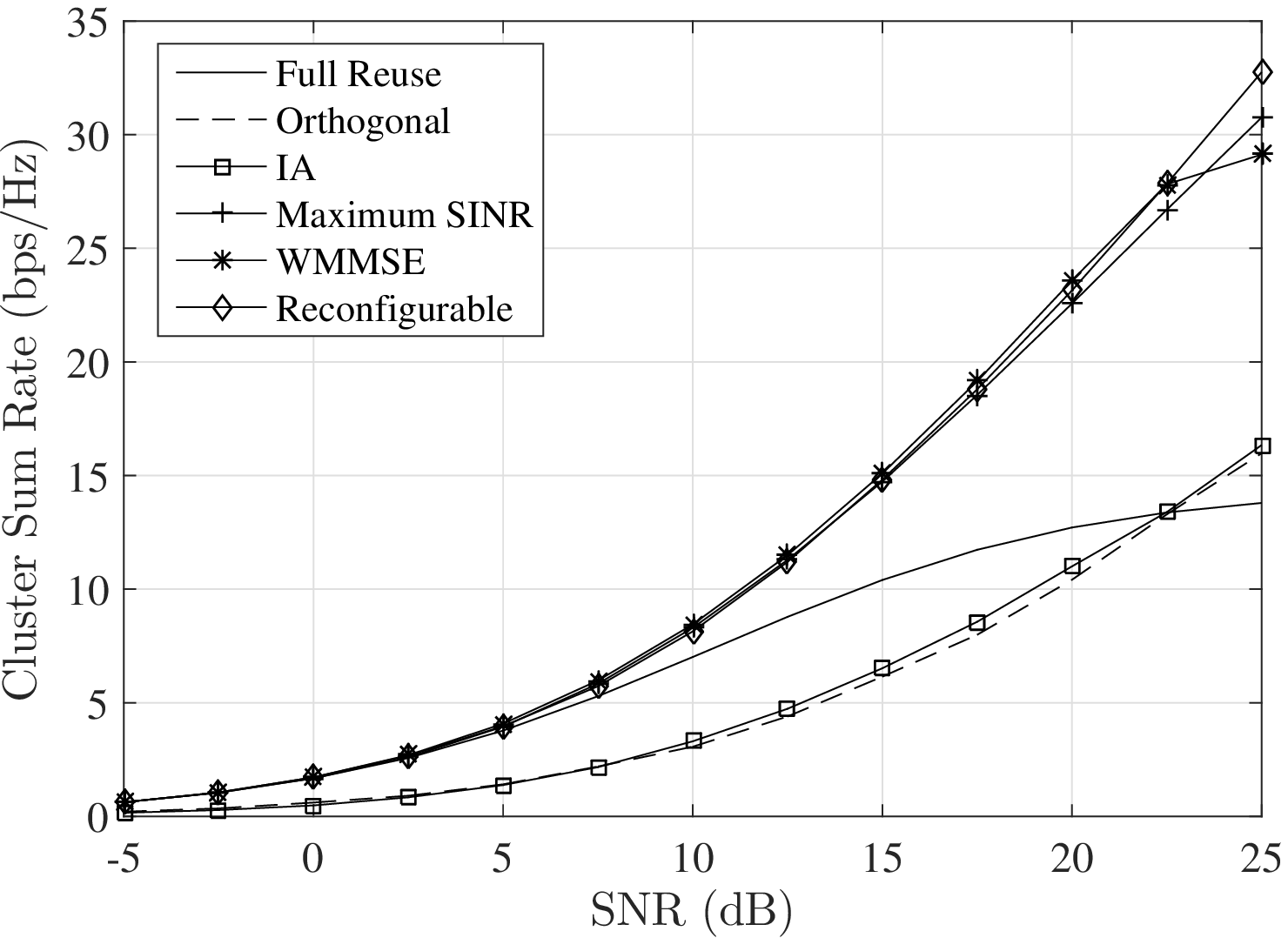} 
\caption{Sum rate performance of various CB schemes as a function of the $\snr$ in dB for $B=3$, $n_{\rm T}=8$, and $n_{\rm R}=4$ as well as $\alpha=1$, $\beta=0$, and perfect CSI. The performances with full resources reuse and orthogonal transmissions are also demonstrated.}
\label{Fig:IFC_Results_1}
\end{figure}
\begin{figure}[t!] 
\centering
\includegraphics[width=3.45in]{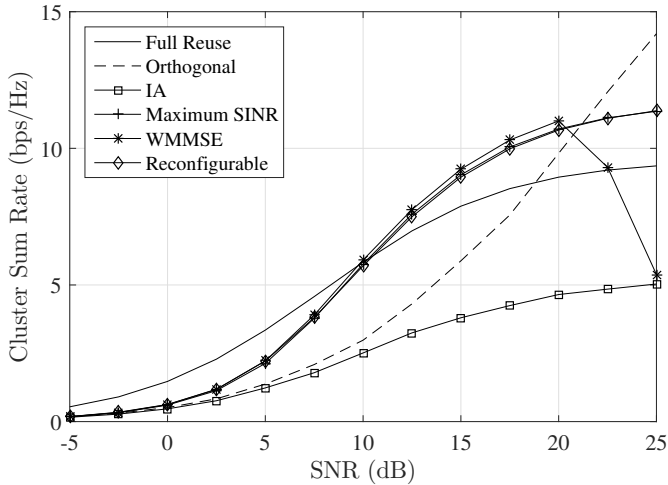} 
\caption{Sum rate performance of various CB schemes as a function of the $\snr$ in dB for $B=3$, $n_{\rm T}=8$, and $n_{\rm R}=4$ as well as $\alpha=0.8$, $\beta=0.2$, and $N_{\rm p}=10$. The performances with full resources reuse and orthogonal transmissions are also demonstrated.}
\label{Fig:IFC_Results_2}
\end{figure}   
\section{Numerical Results and Discussion}\label{sec:Results}
\begin{figure}[t!] 
\centering
\includegraphics[width=3.45in]{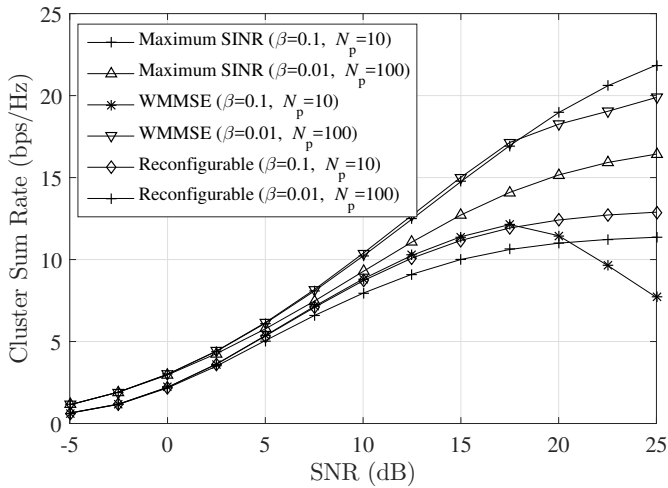} 
\caption{Sum rate performance of the considered iterative CB schemes as a function of the $\snr$ in dB for $B=4$, $n_{\rm T}=4$, and $n_{\rm R}=2$ as well as $\alpha=0.9$, $\beta=\{0.1,0.01\}$, and $N_{\rm p}=\{10,100\}$.}
\label{Fig:IFC_Results_3}
\end{figure}

The average cluster sum rate with the CB schemes described in Section~\ref{sec:Full_Coordination} is assessed hereinafter by simulating \eqref{Eq:Ergodic_SumRate} over $100$ independent channel realizations using the system model introduced in Section~\ref{sec:Cellular_Model}. In Figs$.$~\ref{Fig:IFC_Results_1} and~\ref{Fig:IFC_Results_2}, a coordination cluster with $B=3$ BSs is considered where we have set $n_{\rm T}^{[b]}=n_{\rm T}=8$ as well as $n_{\rm R}^{[b]}=n_{\rm R}=4$ $\forall$$b=1,2,$ and $3$, while a cluster with $B=4$ BSs is investigated in Fig$.$~\ref{Fig:IFC_Results_3}, where $n_{\rm T}=4$ and $n_{\rm R}=2$. For IA in Figs$.$~\ref{Fig:IFC_Results_1} and~\ref{Fig:IFC_Results_2}, $d_b$ for each MT $b$ was set to $0.5\min(8,4)=2$ according to the IA feasibility conditions \cite{J:Jafar_interference} and the precoding matrices $\mathbf{V}_b$'s were obtained in closed form. The latter values for $d_b$'s were also preset to the maximum SINR scheme in Figs$.$~\ref{Fig:IFC_Results_1}--\ref{Fig:IFC_Results_3}, which designs each $\mathbf{V}_b$ iteratively. It is noted that for the system set up in Fig$.$~\ref{Fig:IFC_Results_3}, closed-form expressions for the $\mathbf{V}_b$'s of IA are not available. For both the iterative schemes WMMSE and Reconfigurable in all figures, each $d_b$ was initialized as $d_b=\min(n_{\rm T},n_{\rm R})$ and obtained at the end of the algorithmic iterations or upon convergence, explicitly for Reconfigurable and implicitly for WMMSE, together with all $\mathbf{V}_b$'s. More specifically, the Reconfigurable scheme outputs $d_b$ to be sent by each coordinated BS $b$ together with their beamforming directions, whereas WMMSE only generates the transmit covariances matrices with possibly some streams set to zero power, and thus unusable. This means that, for the latter scheme the optimum $d_b$ needs to be searched in some way, a fact that will cause an extra overhead in practical networks.

As clearly depicted in Figs$.$~\ref{Fig:IFC_Results_1}--\ref{Fig:IFC_Results_3} for a maximum of $10$ iterations per iterative scheme, the performance of all CB schemes is susceptible to high OCI and severe CSI errors. This behavior of the IA  scheme for CSI errors was also observed in \cite{Mungara2015}. For example, for $\snr=15$ dB and $B=3$, it is shown in Figs$.$~\ref{Fig:IFC_Results_1} and~\ref{Fig:IFC_Results_2} that the performance of all CB schemes drops approximately $40\%$ between the two OCI and CSI scenarios, according to which $\alpha$ decreases from $1$ to $0.8$, $\beta$ increases from $0$ to $0.2$, and $N_{\rm p}$ falls to $10$ from a very large number. Interestingly, the maximum SINR, WMMSE, and Reconfigurable schemes that take OCI under consideration provide for most of the $\snr$ values equal to or more than $100\%$ improvement compared to IA. As also seen from Fig$.$~\ref{Fig:IFC_Results_2}, all iterative schemes outperform orthogonal transmission for $\snr\leq20$ dB, and WMMSE faces convergence problems at high $\snr$ values for the considered number of algorithmic iterations. In addition, it is obvious from this figure that, severe CSI errors for $\snr\leq10$ dB render full reuse the best strategy. The trend of Figs$.$~\ref{Fig:IFC_Results_1}--\ref{Fig:IFC_Results_2} is also seen in Fig$.$~\ref{Fig:IFC_Results_3} for a larger cluster size. As depicted, between the WMMSE and Reconfigurable schemes that provide the best performance, the Reconfigurable is more resilient to increasing $\beta$ and decreasing $N_{\rm p}$. At $\snr=15$ dB both schemes differ nearly $20\%$ between the two considered scenarios, whereas at $\snr=22.5$ dB, the Reconfigurable scheme differs $40\%$ while the WMMSE scheme differs closely to $50\%$.

\section{Conclusion}\label{sec:Conclusions}
The feature of coordinating BS transmissions to manage interference in cellular networks is already a part of the latest LTE release offering significant potential for performance improvement especially at the cell edges. From the various recently available CB schemes for cellular MIMO systems, there exist schemes that adapt satisfactory to uncoordinated interference, but still perform poorly under severe CSI imperfections. Certain advances need to take place in order to maximize the benefit from CB, such as for example efficient BS clustering as well as coordination with reduced overhead and resilience to CSI uncertainties. 

\bibliographystyle{IEEEtran}
\bibliography{IEEEabrv,references}
\end{document}